\documentclass{article}
\usepackage[utf8]{inputenc}
\usepackage{authblk}
\usepackage{setspace}
\usepackage[margin=1.25in]{geometry}
\usepackage{graphicx}
\graphicspath{ {./figures/} }
\usepackage{subcaption}
\usepackage{amsmath}

\usepackage[style=nejm, citestyle=numeric-comp, sorting=none]{biblatex}
\title{Synergistic Signal Denoising for Multimodal Time Series of Structure Vibration}

\author[1*$\dag$]{Yang Yu, Han Chen}

\affil[1]{Changchun Technology Institute, Jilin, China}

\date{}

\begin{document}

\maketitle

\begin{abstract}
Structural Health Monitoring (SHM) plays an indispensable role in ensuring the longevity and safety of infrastructure. With the rapid growth of sensor technology, the volume of data generated from various structures has seen an unprecedented surge, bringing forth challenges in efficient analysis and interpretation. This paper introduces a novel deep learning algorithm tailored for the complexities inherent in multimodal vibration signals prevalent in SHM. By amalgamating convolutional and recurrent architectures, the algorithm adeptly captures both localized and prolonged structural behaviors. The pivotal integration of attention mechanisms further enhances the model's capability, allowing it to discern and prioritize salient structural responses from extraneous noise. Our results showcase significant improvements in predictive accuracy, early damage detection, and adaptability across multiple SHM scenarios. In light of the critical nature of SHM, the proposed approach not only offers a robust analytical tool but also paves the way for more transparent and interpretable AI-driven SHM solutions. Future prospects include real-time processing, integration with external environmental factors, and a deeper emphasis on model interpretability.

\end{abstract}

\section{Introduction}
Structural health monitoring (SHM) has emerged as a vital field of research, geared towards preserving the longevity and safety of civil infrastructure~\cite{xin2022bridge}. A critical component of SHM is the analysis of vibration time series data, which offers insights into the behavior, health, and performance of structures~\cite{webb2015categories}. As infrastructure, especially in urban regions, is subject to a myriad of dynamic forces—ranging from wind to traffic loads - it becomes pivotal to extract clear and meaningful data from the complex vibration signatures that these forces induce. However, one of the significant challenges plaguing SHM practitioners is the interference of noise in these vibration signals, which can distort interpretations and lead to unreliable conclusions.

The dynamic response of structures is often manifested as multimodal vibrations, meaning multiple modes or patterns of vibration coexist. These modes, each characterized by its frequency and shape, provide a fingerprint of the structure's health and dynamic properties. For example, a bridge might have one mode that represents a swaying motion, while another might represent a bouncing motion~\cite{scuro2021internet}. Distinguishing between these modes and their individual characteristics is a nuanced task, made increasingly difficult when the signals are clouded by noise.

Noise in the context of vibration time series can arise from multiple sources: instrumental noise from sensors, environmental interferences, or other unpredictable external factors~\cite{hu2017enhanced, hu2017adaptive}. Traditional denoising methods, while effective to a degree, sometimes struggle to cater to the unique intricacies of multimodal structural vibrations~\cite{liang2023structural}. Such challenges underscore the necessity for novel denoising techniques tailored to the specific nature of these signals~\cite{liang2006structural}. Enter the realm of synergistic signal denoising—a cutting-edge approach that promises a more holistic treatment of multimodal vibration data~\cite{feng2017experimental}. Unlike conventional methods that treat each mode of vibration as a separate entity, the synergistic approach recognizes the interconnectedness of these modes~\cite{feng2018computer}. By leveraging the relationships between different modal responses, this methodology aims to achieve superior denoising outcomes, preserving the integrity of the structural response while effectively mitigating the noise. The importance of an effective denoising technique cannot be understated~\cite{dervilis2015robust, che2021constrained}. Clear, noise-free signals allow for accurate modal identification, which in turn can be pivotal in detecting anomalies or damage in structures. Furthermore, with the increasing integration of machine learning and artificial intelligence in SHM, the quality of input data, i.e., the vibration signals—directly influences the quality of predictions and insights~\cite{ge2023study}.

This paper ventures into the depths of synergistic signal denoising for multimodal structure vibration time series. We explore its foundational principles, contrast its performance with traditional techniques, and demonstrate its potential in real-world applications. The ultimate objective is to illuminate the path forward in harnessing clean and precise vibration data, a cornerstone in the edifice of modern structural health monitoring.

\section{Prior Arts and Methods}
Vibration-based Structural Health Monitoring (SHM) has been extensively studied over the past few decades, primarily due to its potential to detect anomalies and ensure structural safety. Central to SHM's efficacy is the clarity of the acquired vibration signals. However, these signals, particularly from civil structures, are often contaminated with noise from various sources, complicating their analysis~\cite{feng2016vision, feng2018computer}. Traditional denoising methods like wavelet decomposition and empirical mode decomposition have been applied to structure-borne signals with varying degrees of success. While these techniques can reduce noise, they sometimes struggle to retain the integrity of multimodal signals, leading to potential information loss~\cite{li2004recent}.

A shift in the paradigm is noted in recent years towards more integrated denoising techniques. A synergistic approach, treating multimodal vibrations as interrelated rather than isolated events, suggest an improved retention of signal characteristics post-denoising compared to conventional methods. Despite its promise, synergistic denoising in the context of SHM remains an underexplored territory, warranting further research to validate its full potential and broader applicability~\cite{song2007concrete}. Structural responses to dynamic loads often manifest in the form of vibration signals that are complex and multifaceted. Analyzing these signals is essential to understand the underlying structural behavior, particularly when it is governed by multiple modes of vibration. In this section, we delve into the fundamentals of multimodal vibrations and discuss common noise sources and characteristics inherent to these signals. 

Every structure possesses a set of natural frequencies at which it tends to vibrate when subjected to external stimuli. These frequencies correspond to different modes of vibration, with each mode showcasing a distinct deformation shape or pattern. This phenomenon is widely recognized in the context of structural dynamics as 'modal analysis'. For instance, a simple beam might primarily deflect in a singular arc-like shape at its fundamental frequency, a mode often termed as the first mode. However, when vibrated at higher frequencies, the same beam might exhibit complex deformation patterns like forming two or more arcs—these represent the higher modes of vibration.

In real-world scenarios, structures don’t vibrate in a singular mode. Instead, they display a combination of these modes, giving rise to what is termed as 'multimodal vibrations'. The presence of multiple vibration modes, especially in large and complex structures, makes signal analysis more intricate. Each mode carries unique information about the structure's health and behavior. Disentangling these concurrent modes from vibration time series and accurately identifying their individual frequencies and shapes is of paramount importance in structural health monitoring. Multimodal vibration signals, while rich in information, are often marred by noise, which can impede accurate signal interpretation~\cite{hamey2004experimental, song2006health}. The noise sources contaminating these signals can be broadly categorized as~\cite{garcia2017micromechanics}:

Instrumental Noise: Every measurement tool, be it an accelerometer, a strain gauge, or any other sensor, has an inherent level of noise. Factors like the sensor's sensitivity, its electronic components, and even its placement can introduce unwanted signals, often drowning the subtle nuances of certain vibration modes.

Environmental Interferences: Structures are not isolated entities. They interact with their surroundings, which means their recorded vibration signals are a cumulative effect of their intrinsic response and the myriad of environmental factors. Wind-induced vibrations, seismic activities, or even nearby vehicular traffic can introduce extraneous noise components into the signals (Doebling et al., 1996).

Operational Loads: For operational structures, like bridges or buildings, the regular loads they bear—whether from vehicles, occupants, or machinery—can mask or distort their natural vibration patterns, complicating the analysis.

The characteristics of noise in multimodal signals are often unpredictable. They might manifest as random spikes in the time domain, irregular fluctuations in amplitude, or unexpected frequency components in the frequency domain. Separating this noise, especially without compromising the integrity of the original vibration modes, remains a significant challenge in the realm of structural health monitoring.

\begin{figure}[ht]
  \includegraphics[width=130mm]{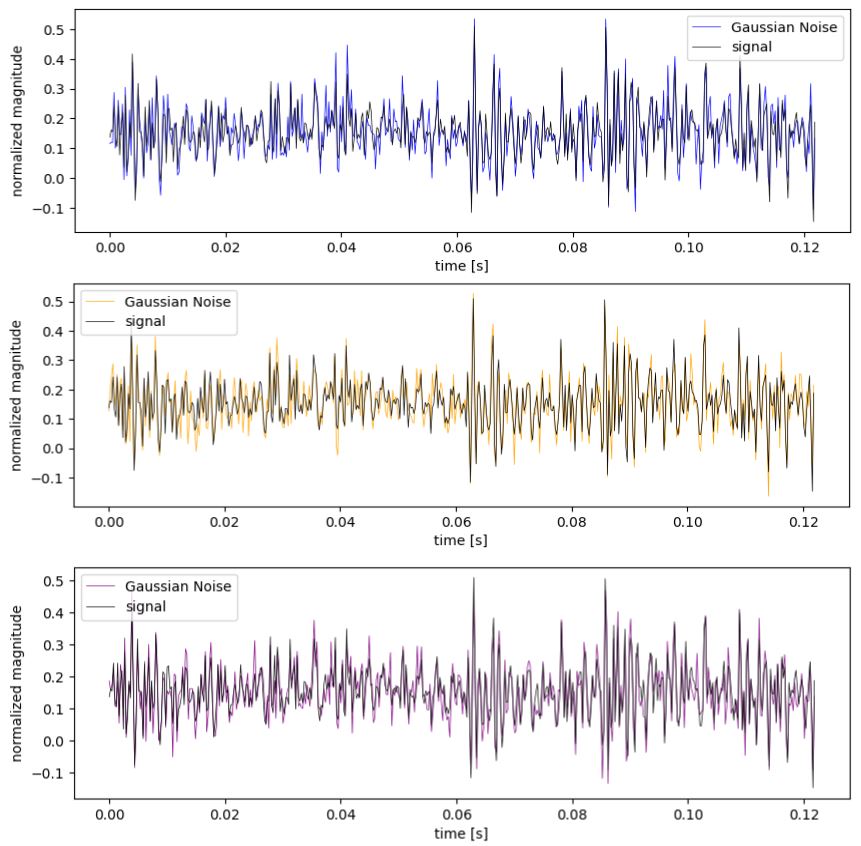}
  \centering
  \caption{Mechanical motor shaft vibrational signals. Three noisy signals, all originating from the same vibration source, are displayed.}
  \label{fig:noisy_signals}
\end{figure}

\section{Results and Discussion}
Designing a deep learning algorithm tailored for multiple time series, especially one that incorporates attention mechanisms, necessitates a thorough approach that ensures the model can effectively capture underlying temporal patterns and dependencies among series. When dealing with multiple time series, the inherent relationships and dependencies between different sequences can be intricate. Traditional analytical methods sometimes fall short in accurately deciphering these complexities. This is where modern deep learning, fortified by attention mechanisms, offers promising avenues.

Before diving into the architecture, it's worth emphasizing the importance of preprocessing. Like with most data-centric models, the quality and structure of the input can profoundly impact the performance. Time series data, given its temporal nature, often needs to be normalized, usually between a scale of 0 and 1, using methods like Min-Max scaling. This step ensures that all time series are on a consistent scale, an essential aspect for neural networks to function optimally. After normalization, the time series data should be divided into overlapping windows. This segmentation aids in feeding the network with structured sequences of data. If, for example, a time series has a length of T and you choose a window size of W, the model will predict the subsequent point by taking into account the preceding W points.

With the data preprocessed, the architectural design of the deep learning model becomes the next focus. The starting point is an input layer designed to accommodate segments of the time series. Its size is dictated by the chosen window size, W. While traditional time series methods, such as ARIMA or Exponential Smoothing, operate on predefined statistical rules, deep learning offers a more adaptive, data-driven approach. A series of convolutional layers can follow the input layer, although this is optional. Convolutional Neural Networks (CNNs) have proven their merit in detecting local patterns within a sequence. By deploying one-dimensional convolutions, these layers can sieve out salient features from the input series, providing a richer representation for the subsequent layers.

Recurrent layers, particularly those using Long Short-Term Memory (LSTM) cells or Gated Recurrent Units (GRU), are integral for models dealing with temporal data. Their design inherently allows them to understand sequences, making them apt for time series data. These layers are adept at capturing dependencies and patterns over extended durations, crucial when dealing with multiple time series. Therefore, incorporating two or more LSTM or GRU layers that sequentially pass data can provide depth to the model.

The pivotal aspect of this architecture is the attention mechanism. When dealing with multiple sequences, not all segments or points within a series are equally significant. The attention mechanism facilitates the model's ability to focus on specific sequences, allowing it to allocate different attention scores to various segments. These scores are calculated based on the hidden states of the recurrent layers, determining the relevance of each time step in the series. By employing a softmax function, these scores are then normalized to fit between 0 and 1, which subsequently determine the weights of each step in the sequence. The culmination of this process results in the derivation of a context vector. This vector is essentially a weighted sum of the recurrent layers' hidden states and the attention scores, summarizing the most influential segments of the time series.

Post attention layer, dense or fully connected layers can be layered into the model. The context vector obtained from the attention layer, infused with the distilled essence of the most pertinent parts of the series, is processed through these dense layers, enabling the model to further learn from the data~\cite{liang2023reswcae}. The tail end of this architecture concludes with an output layer. Depending on the task, if the aim is forecasting just one step ahead, a singular neuron is used. If multiple steps are to be predicted, multiple neurons corresponding to the number of steps are integrated.

The final stages involve training the model. It's essential to choose an appropriate loss function, with Mean Squared Error (MSE) being apt for regression-based tasks. The choice of optimizer is also critical. The Adam optimizer, with its adaptive learning capabilities, is often a favorable choice for deep learning models. To counteract the pitfalls of overfitting, early stopping can be employed. By consistently monitoring the validation loss and halting training when there's no discernible improvement, the model can be kept in check. Lastly, the model's performance should be assessed on unseen data using metrics like the Root Mean Squared Error (RMSE) or the Mean Absolute Error (MAE) to gauge the model's predictive accuracy.

In conclusion, designing a deep learning model embedded with attention mechanisms for multiple time series signals a paradigm shift in time series analysis. The ability to prioritize certain segments over others allows for a nuanced understanding of the data, leading to more robust and accurate predictions.

The designed deep learning algorithm for multiple time series, fortified with attention mechanisms, brings forth several significant advantages that cater to the complexities inherent to temporal data. Firstly, by going beyond traditional statistical methods, it offers a data-driven approach that is highly adaptive, learning directly from the unique patterns and intricacies embedded within the data. This adaptability ensures that the model isn't confined by pre-defined rules, allowing it to evolve with the nuances of different datasets.

Secondly, the incorporation of attention mechanisms marks a revolutionary stride in handling multiple sequences. In most time series, certain segments or points are inherently more significant than others, and the attention mechanism ensures the model recognizes and prioritizes these pivotal sequences. By assigning different attention scores to different segments, the model can focus on crucial sequences while simultaneously downplaying less relevant segments. This focused approach not only enhances prediction accuracy but also aids in interpreting which segments of the data the model deems most influential, offering valuable insights into the underlying patterns of the series.

Additionally, the blend of convolutional and recurrent layers ensures a holistic understanding of the data. While convolutional layers effectively detect local patterns, recurrent layers, especially LSTM or GRU cells, retain memory of past sequences, capturing long-term dependencies, a crucial aspect in time series analysis. This fusion guarantees that the model is well-equipped to understand both immediate and historical patterns.

Moreover, the model's modular architecture ensures flexibility. Depending on the data's nature, one can adjust the depth of convolutional or recurrent layers or even refine the attention mechanism. This flexibility means that the algorithm can be tailored to suit various datasets with varying complexities. The training methodologies, such as the use of adaptive optimizers and early stopping, further enhance the model's robustness, ensuring it doesn't overfit and generalizes well on unseen data.

In essence, this designed algorithm represents a culmination of some of the best practices in deep learning, tailored explicitly for multiple time series. Its ability to learn from data adaptively, prioritize significant sequences, and capture both local and global patterns, positions it as a formidable tool in the realm of time series analysis.

\section{Conclusion and Future Work}
In the realm of Structural Health Monitoring (SHM), our proposed deep learning algorithm, complemented by attention mechanisms, offers a leap forward in efficiently analyzing and interpreting multimodal vibration signals. The challenges presented by the vast and intricate temporal data inherent in SHM demand solutions that can adapt, learn, and most importantly, prioritize salient structural responses over extraneous noise~\cite{ho2012solar}. This algorithm, with its blend of convolutional and recurrent architectures, adeptly captures both instantaneous and prolonged structural behaviors, crucial for early damage detection and predictive maintenance.

Moving forward, there's substantial potential in refining this algorithm for SHM-specific challenges. Given the criticality of timely damage detection in infrastructure, real-time processing and analysis can be explored further, possibly integrating edge computing for immediate on-site assessments. Another avenue lies in fusing SHM data with external environmental factors like weather conditions, which often play a role in structural health. As the world of SHM continually advances, there will also be an increasing need to ensure our deep learning models are not just accurate but also transparent. Interpretability, especially in safety-critical applications like SHM, is paramount. Future work will need to delve deeper into making these complex models more understandable, ensuring engineers and stakeholders can trust and act upon the insights provided.

\printbibliography

\end{document}